\documentclass[11pt]{article}
\usepackage{cite}
\usepackage{amssymb,amsmath,amsthm,mathrsfs}

\usepackage[english]{babel}
\usepackage{graphicx}
\usepackage{dcolumn}
\usepackage{bm}
\usepackage{color}

\usepackage{hyperref}

\usepackage{subfigure}
\usepackage[ansinew]{inputenc}
\usepackage{amsfonts}
\usepackage{amsmath} 
\def\<{\langle}
\def\>{\rangle}
\def\bra#1{\langle #1 |}
\def\ket#1{| #1 \rangle}

\def\e{\mathrm{e}}

\def\e{e}
\def\i{i}

\newcommand{\As}{\mathcal{A}}

\def\bra#1{\langle #1 |}
\def\ket#1{| #1 \rangle}

\def\Ord{\mathrm{O}}

\newcommand{\beq}{\begin{equation}}
\newcommand{\eeq}{\end{equation}}
\newcommand{\barr}{\begin{eqnarray}}
\newcommand{\earr}{\end{eqnarray}}

\def\e{\mathrm{e}}

\begin{document}

\title{\bf The Legacy of George Sudarshan }
\author{Giuseppe Marmo \\
Dipartimento di Fisica and MECENAS, Universit\`a di Napoli, I-80126 Napoli, Italy \\
INFN, Sezione di Napoli, I-80126 Napoli, Italy 
\\ [1ex]
email: giuseppe.marmo@na.infn.it \\ \\
Saverio Pascazio \\
Dipartimento di Fisica and MECENAS, Universit\`{a} di Bari, I-70126 Bari, Italy \\
INFN, Sezione di Bari, I-70126 Bari, Italy \\
INO-CNR, I-50125 Firenze, Italy \\
[1ex]
e-mail: saverio.pascazio@ba.infn.it}

\maketitle

\section{Ennackal Chandy George Sudarshan}
\label{ecgs}

Ennackal Chandy George Sudarshan was born on 16 September 1931 in Pallam, a small village in the Kottayam District of Kerala State, South India. He was raised in a Christian family, although he later considered himself closer to the Hindu religion. His mother, Achamma, was a school teacher; his father, E.\ I.\ Chandy, was a revenue supervisor for the Kerala Government Service.

After obtaining his master's degree at the University of Madras, in 1951 he moved to the Tata Institute of Fundamental Research in Bombay (Mumbai), directed by Homi Bhabha. The presence of a strong group in mathematics allowed him to become acquainted with topology, modern algebra, the theory of spinors, and  functional analysis. There, he started by collaborating with B.\ Peters on cosmic rays research, and quickly made one of his first discoveries: he realized that he could be a theoretical physicist, inspired by experimental research, rather than become an experimental physicist.  He then moved to the University of Rochester in New York, where he got his PhD under the supervision of Robert Marshak. Subsequently, he went to work at Harvard University with Julian Schwinger.
In 1959, he went back to Rochester as an Assistant Professor, and two years later became Associate Professor.
Five years later he moved to Syracuse University as a Professor, where he gave birth to a group in elementary particle physics. Up to then the major emphasis at Syracuse had been on general relativity, the group being 
led by Peter G. Bergman. In 1969, Sudarshan made his last move from Syracuse to the University of Texas at Austin, as a Professor and Co-Director (with Yuval Ne'eman) of the Centre for Particle Theory. He would later convince Steven Weinberg to move to Austin.

George was an eclectic thinker. It is just plain impossible to make an exhaustive list of the subjects in which he was involved and gave seminal contributions. In theoretical physics, he pioneered the V-A theory of weak interactions \cite{V-A}, the role of symmetry in high-energy physics \cite{MOS,OMS}, the spin-statistics theorem \cite{spinstat}, quantum optics, including the theory of optical coherence \cite{qopt,KS}, the quantum Zeno effect \cite{MS}, the hypothesis of tachyons and faster than light propagation \cite{BDS,AS}, the no-interaction theorem for classical relativistic Hamiltonian theories of point particles \cite{nointth,nointthbook},
the theory of open quantum systems, from quantum dynamical semigroups \cite{GKS} to quantum channels \cite{SMR}, and quantum applications, from quantum computing to tomography \cite{qtomo,qtomo2}.
Outside physics, he gave contributions on the relations between East and West, as well as politics, history of science, philosophy and religion.

In the following we shall briefly summarize George's contributions in the area of quantum evolutions, starting from the quantum Zeno effect and ending up with quantum channels and quantum semigroups. This is but a small part of his oeuvre. Yet, as we will see, these contributions are of crucial importance in the booming field of quantum mechanics and applications. We find it curious that one of the main applications of the quantum Zeno effect consists in the control of decoherence and dissipation; in turn, these are just the topics pioneered by George Sudarshan in the context of open quantum systems (Gorini-Kossakowski-Lindblad-Sudarshan equation and Kraus-Sudarshan representation).

\section{Quantum time evolutions}
\label{tut}

Quantum evolution is described by the Schr\"odinger equation (henceforth $\hbar = 1$)
\begin{equation}
\label{schr}
\i \dot \psi \;=\; H \psi \,,
\end{equation}
where $\psi$ is the wave function, the dot denotes time derivative, and $H$ is the Hamiltonian of the system.
The formal solution is
\begin{equation}
\label{schrt}
\psi_t \;=\; U_t \psi_0 \,,
\end{equation}
where $U_t= \e^{-\i H t}$ and $\psi_0$ is the initial condition.  In terms of the density matrix $\rho$, the evolution is governed by the von Neumann equation
\begin{equation}
\label{vn}
\dot \rho \;=\; -\i [H,\rho] \,,
\end{equation}
whose solution with initial condition $\rho_0$ is 
\begin{equation}
\label{vnt}
\rho_t \;=\; U_t \rho_0 U^\dagger_t\,.
\end{equation}
Here and in the following we shall not linger on mathematical rigor (boundedness of operators, normalizability of states, existence of moments, and so on) and proper definitions, but rather focus on George Sudarshan's guiding ideas. We shall  review in this Section three of his seminal contributions in the area of quantum dynamics.

\subsection{Quantum Zeno}
\label{zeno}

The survival amplitude and probability in the initial state $\psi_0$ are
\begin{eqnarray}
\label{survamp}
\As(t) &=& \bra{\psi_0} U_t \ket{\psi_0},  \\
\label{survprob}
p(t) &=& |\As(t)|^2, 
\end{eqnarray}
respectively. A short time expansion reads
\begin{eqnarray}
\label{survpst}
p(t) &=& 1 - \left(\frac{t}{\tau_Z}\right)^2 + \Ord(t^4) \\
\label{zt}
& & \tau_Z = \left( \bra{\psi_0} H^2 \ket{\psi_0} - \bra{\psi_0} H \ket{\psi_0}^2 \right)^{-1/2}.
\end{eqnarray}   
The survival probability is thus quadratic at short times, and its curvature at the origin is the ``Zeno" time $\tau_Z$. The above conclusion (\ref{survpst}) is very general, and valid also for ``unstable" states, $p(t)\to 0$ (with an approximately exponential law) when $t\to\infty$. The quadratic behaviour leads to the quantum Zeno effect: if the system undergoes frequent measurements, to check whether it is still in its initial state, it does not evolve, as 
\begin{equation}
\label{QZ}
[p(t/N)]^N \stackrel{N \to \infty}{\longrightarrow}  1 \,.
\end{equation} 
The (elegant) allusion to the sophist philosopher is due to Misra and Sudarshan \cite{MS}. Incidentally, Zeno was born in Velia (Elea in ancient Magna Graecia), not far from the birthplace of one of the authors of this manuscript \cite{srufo}.
The fecund idea (\ref{QZ}) motivated many experiments and led to many applications in quantum information, quantum technologies, and quantum control.

\subsection{Sudarshan-Mathews-Rau (and the Kraus-Sudarshan representation)}
\label{SMR}

If the system $S$ under consideration interacts with a larger system $R$, the dynamics is still governed by Eqs.\ (\ref{schr}) and (\ref{vn}), written in the form
\begin{equation}
\label{schrtot}
\i \dot \psi_{\textrm{tot}} \;=\; H_{\textrm{tot}} \psi_{\textrm{tot}} \quad \longleftrightarrow \quad
 \dot \rho_{\textrm{tot}} \;=\; -\i [H_{\textrm{tot}},\rho_{\textrm{tot}}] \,,
\end{equation}
with $\psi_{\textrm{tot}}, \rho_{\textrm{tot}}$ and $H_{\textrm{tot}}$ the wave function, density matrix and Hamiltonian of the total system $R+S$, respectively.
System $S$ is not isolated anymore and must be considered an \textit{open} quantum system. It was far from being obvious, in 1961, that the dynamics of open quantum systems could be characterized in some way. Sudarshan, Mathews, and Rau published an article \cite{SMR} showing that the density matrix of system $S$
\begin{equation}
\label{denss}
\rho \equiv \rho_S = \textrm{Tr}_R \, \rho_{\textrm{tot}}
\end{equation}
must obey the evolution
\begin{equation}
\label{superoperator}
\rho' \;=\; \Lambda \rho \;=\; \sum_\alpha W_\alpha  \rho W_\alpha^\dagger ,
\end{equation}
where $W_\alpha$'s are operators on the Hilbert space of system $S$, and $\sum_\alpha W_\alpha^\dagger  W_\alpha = \mathbb{I}$, the identity operator. Notice that if there is single operator $W = U = \e^{-\i H t}$ in the above summation, Eq.\ (\ref{superoperator}) reduces to Eq.\ (\ref{vnt}): in modern language, we would say that system $S$ undergoes no decoherence.
Equation (\ref{superoperator}) is known nowadays as a quantum channel, and $\Lambda$ is named superoperator. 
In fact, Eq.\ (\ref{superoperator}) is often called the Kraus representation of a quantum channel. Sudarshan, Mathews, and Rau derived it almost ten years before Kraus \cite{G5,KRAUS}. Their proof did not make use of the notion of complete positivity and contained a gap \cite{CP2017}, as the evolution matrix they considered, being re-aligned \cite{Karol}, is block-positive (and not necessarily positive). 
Quantum channels and their Kraus-Sudarshan representation (\ref{superoperator}) are nowadays routinely used in innumerous quantum applications.

\subsection{Gorini, Kossakowski and Sudarshan (and Lindblad)}
\label{GKLSequation}

The evolution law (\ref{superoperator}) represents a ``snapshot" of the density matrix at a given time $t^*$: $\rho' = \rho_{t^*}$. Can one describe the system dynamics in terms of a differential equation, free from memory effects? This is a difficult question. Equations of this type were known as master or kinetic equations and had been considered by Landau as early as 1927 \cite{Landau}. 
Gorini, Kossakowski and Sudarshan \cite{GKS} (and independently Lindblad \cite{Lindblad}) solved the problem by writing the following differential (``master") equation
\begin{equation}
\label{master}
\dot \rho \;=\;  \mathcal{L}\rho  
\end{equation}
and characterizing the generator $\mathcal{L}$ in terms of quantum dynamical semigroups. 
The explicit expression is \cite{GKS} 
\begin{equation}
\label{GKS}
 \dot \rho \;=\;   \mathcal{L} \rho \;=\; -\i [H,\rho] + \frac 12 \sum_{k,l} c_{kl} \Big( [F_k,\rho F_l^\dagger] +  [F_k\rho, F_l^\dagger]  \Big) \,,
\end{equation}
or equivalently \cite{Lindblad}
\begin{equation}
\label{GKSL-2}
 \dot \rho \;=\;   \mathcal{L}\rho \;=\;  -\i [H,\rho] + \frac 12 \sum_j (2V_j  \rho V_j^\dagger - V_j^\dagger V_j \rho - \rho V_j^\dagger V_j)\, .
\end{equation}
In both cases $H$ is self-adjoint and represents the Hamiltonian of the system. The operators $F_k$ and $V_j$ act on the system Hilbert space, and $[c_{kl}]$ is a complex positive matrix (Kossakowski matrix \cite{Kossa1972}). Equations (\ref{GKS}) and (\ref{GKSL-2}) are completely equivalent and are known as Gorini, Kossakowski, Lindblad and Sudarshan (GKLS) equation. 

One needs several physical assumptions (among them, the Markovian approximation) to derive Eqs.\ (\ref{GKS})-(\ref{GKSL-2}). 
The GKLS equation is of tantamount importance in the description of open quantum systems, Markovian quantum channels, quantum information, quantum communication and quantum applications.

\section{Anecdotes}
\label{anecd}

\begin{figure}
\centering
\includegraphics[width=120mm]{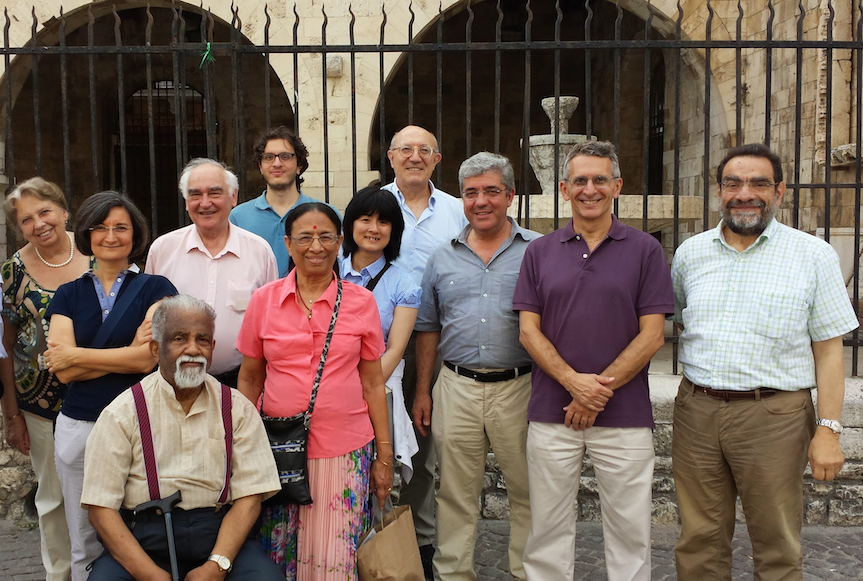}
\caption{\small From left to right: M.\ Man'ko, E.\ Ercolessi, G.\ Sudarshan, V.\ Man'ko, L.\ Ferro, 
Bhamathi, C.\ Uchiyama, F.\ Ventriglia, G.\ Capriati, S. Pascazio, and G. Marmo, old town, Bari. }
\label{fig:group}
\end{figure}

George was endowed with a razor-sharp wit and a brilliant personality. His comments would often come out of the blue and leave his interlocutors stunned. We like to remember some of his jokes and comments. There are far too many, so we will opt for a personal choice.

In 1999 P.\ Facchi and S.\ Pascazio went to Naples, where G. Marmo was hosting George and Vittorio Gorini. The discussion hinged on the conditions that would yield the quantum Zeno effect. At the blackboard, someone said: the Hamiltonian is self-adjoint. After a split second George replied: why should a Hamiltonian be self-adjoint? Are symmetric Hamiltonians not sufficient? An article was born after that discussion \cite{FGMPS}, but the question still resounds in our memory. George would never take a postulate for granted.

A few years later, always in Naples (George was a regular visitor in Italy), while eating a delicious Neapolitan pizza, some of us were having an animated discussion  on quantum mechanics and the projection postulate. George would listen, silent. At some point he said: it is a good thing that quantum mechanics does not depend on its foundations \cite{Gquote}.

Years ago, George and his wife Bhamathi were flying from India back to Texas. Their flight was badly delayed and they missed their connecting flight, somewhere in the US. Bhamathi and George were exhausted after the long trip, and George went to the counter of Americal Airlines, trying to buy a ticket to Austin, Texas. At the AA counter he was asked for an outrageous price. He replied right away: I said I wanted to buy a ticket, not the whole aircraft.

On 3 July 2014 George was invited to give a public talk at the Physics Department of the University of Bari. The lecture was on Weak Interactions and the auditorium was full. There were many questions at the end of the seminar. Someone (either a student or a postdoc) asked George an opinion about the discovery of the Higgs particle (that had been officially announced at CERN two years before). George said: young man, I would not worry about Higgs, I would rather ask: why the muon?

\section{The Legacy of George Sudarshan }
\label{legacy}

George made a number of profound discoveries. His inheritance is so vast that it is difficult to gauge it. 

George Sudarshan was awarded the 2010 Dirac Medal together with Nicola Cabibbo, ``in recognition of their fundamental contributions to the understanding of weak interactions and other aspects of theoretical physics." 
The story of the V-A theory is very well told by Sheldon Glashow \cite{Glashow} (boldface in the original):
``Marshak and Sudarshan, having pondered these matters for some months, met with Felix Boehm in July, 1957, where they learned that the latest data disfavored the VT possibility that Schwinger had adopted. They soon were able to complete and submit their ground-breaking paper, ``The Naure of The Four-Fermion Interaction," of which Sudarshan (then a mere graduate student) was the first-named author. They presented a comprehensive analysis of the weak interaction data, which along with the imposition of an elegant symmetry principle, allowed them to deduce a unique form for the weak interactions, the so-called V-A theory. Their daring hypothesis was accompanied by a list of four experimental results that, they wrote, ``cannot be reconciled with this hypothesis... All of these experiments should be redone... If any of the four experiments stand, it will be necessary to abandon the hypothesis." \textbf{This is theoretical physics at its zenith!} The experiments were redone with results that now confirmed their hypothesis. It was a stunning accomplishment, yet one which has never been recognized with a prize."
Richard Feynman acknowledged George Sudarshan's contribution in 1963 stating that the ``V-A theory was invented by Marshak and Sudarshan, published by Feynman and Gell-Mann, and completed by Cabibbo" \cite{Glashow,Mehra}. Steven Weinberg stressed that V-A was the key \cite{Weinberg}.

The Sudarshan-Glauber representation of coherent light, for which Roy Glauber was awarded the 2005 Nobel prize in Physics, is a cornerstone in quantum optics. The GKLS equation, the Kraus-Sudarshan representation, the quantum Zeno effect are other remarkable contributions, whose (largely unsuspected) importance and relevance have boomed with the advent of quantum applications. 

George is one of the greatest scientist India has ever given birth to. He stands with Ramanujan, Raman, Bose and Chandrasekhar.
His scientific work combines Western logic and precision with Eastern imagination; Greek philosophy with an ancient Indian way of thinking.
George was an iconoclast, with a marked absence of all-pervading cultural prejudices: who else could have thought of what we nowadays call tachyons?

His seminal work is repeatedly quoted but is often unread, and is generally not really understood until considerable time has passed. The deep common thread of his contributions has been the attempt to resolve the fundamental conflict between ``being" and ``becoming", objects and processes. This is one central problem in natural philosophy: to understand and describe the meaning of existence and the nature of change.

The three pictures show George Sudarshan and some of his collaborators, including the authors of this note. George left us on 14 May 2018. His cleverness, humor, keenness of intellect and unparalleled swift of thoughts will be deeply missed.

\begin{figure}
\centering
\includegraphics[width=90mm]{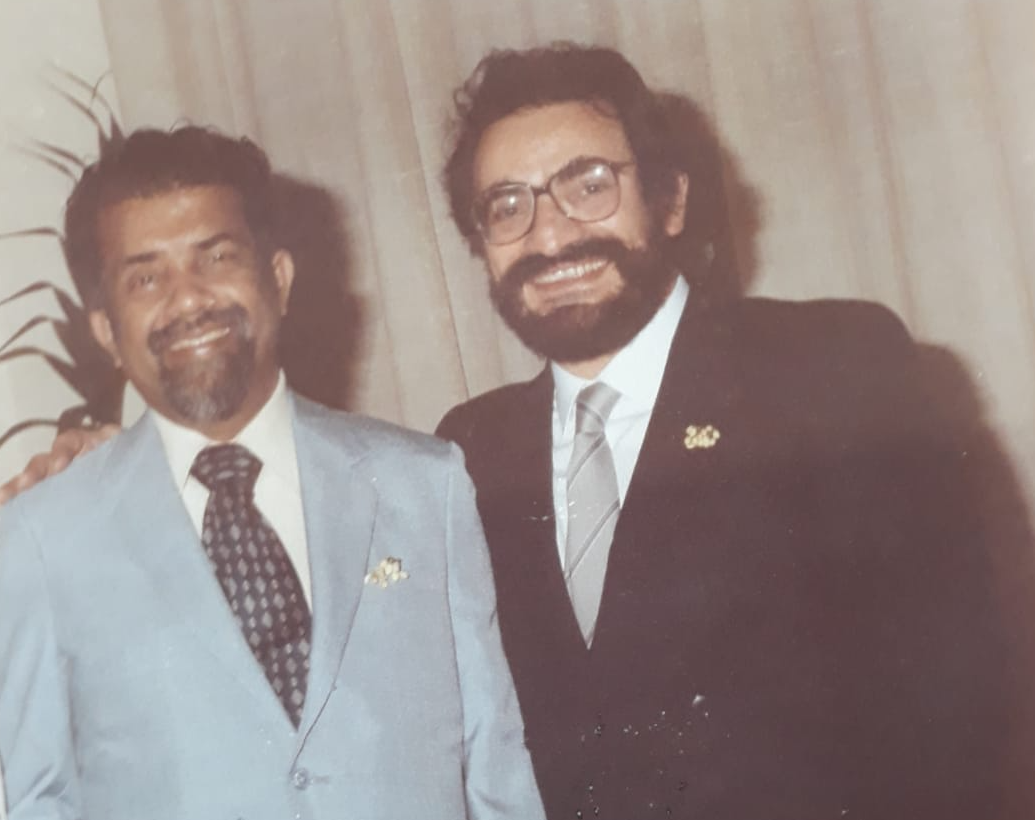}
\caption{\small George Sudarshan and one of the authors at a wedding.}
\label{fig:beppe_george}
\end{figure}
\begin{figure}
\centering
\includegraphics[width=120mm]{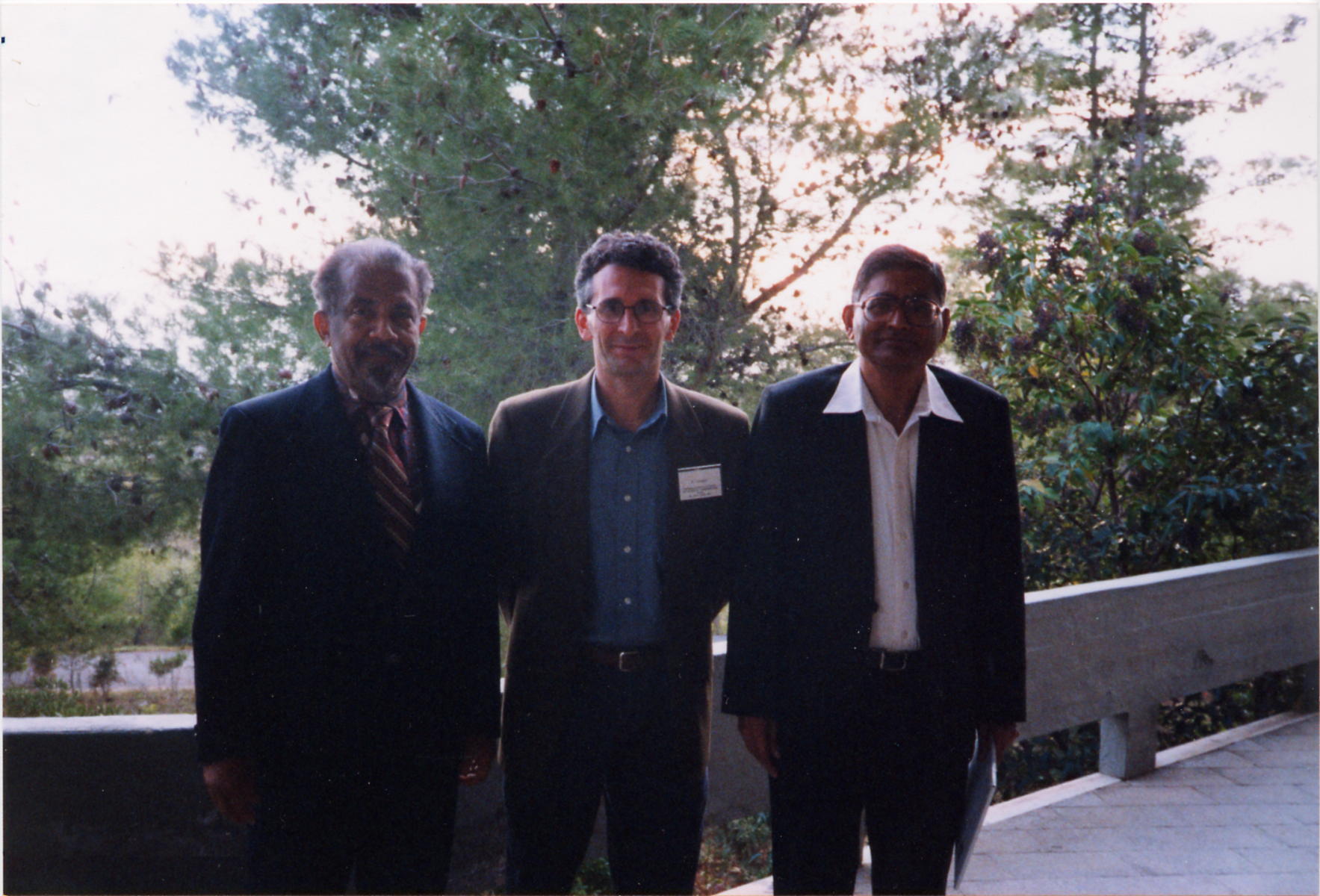}
\caption{\small Applied quantum Zeno effect at a Solvay conference \cite{solvay}: one of the authors of this manuscript cannot move, as he proudly stands at the center of the Misra-Sudarshan theorem.}
\label{fig:MS}
\end{figure}

\section*{Acknowledgments}
G.\, M.\, would like to thank partial financial support provided by the Santander/UC3M Excellence Chair Program 2019/2020.
S.\,P.\, is partially supported by INFN through the project ``QUANTUM".

\end{document}